\documentclass[useAMS,usenatbib]{mn2e}

\usepackage{epsfig}
\usepackage{amsmath}
\usepackage{amssymb}
\usepackage{natbib}
\usepackage{threeparttable} 
\usepackage{booktabs}
\usepackage{color}
\usepackage{multirow}
\usepackage{hyperref}
\usepackage{gensymb}

\usepackage{epstopdf}
\usepackage{times}

\def \aj {AJ}
\def \mnras {MNRAS}
\def \apj {ApJ}
\def \apjs {ApJS}
\def \apjl {ApJL}
\def \aap {A\&A}
\def \nat {Nature}

\title[Radio variability in Sw J1858.6-0814]{The variable radio counterpart of Swift J1858.6-0814}
\author[van den Eijnden et al.]
{J. van den Eijnden$^{1}$\thanks{e-mail: A.J.vandenEijnden@uva.nl}, 
N. Degenaar$^{1}$, 
T. D. Russell$^{1}$,  
D. J. K. Buisson$^{2}$,
D. Altamirano$^{2}$, 
\newauthor M. Armas Padilla$^{3,4}$,
A. Bahramian$^{5,6}$, 
N. Castro Segura$^{2}$, 
F. A. Fogantini$^{7,8}$, 
C. O Heinke$^{9}$, 
\newauthor T. Maccarone$^{10}$, 
D. Maitra$^{11}$, 
J. C. A. Miller-Jones$^{6}$, 
T. Mu\~noz-Darias$^{3,4}$, 
\newauthor M. \"Ozbey Arabac{\i}$^{2,12}$, 
D. M. Russell$^{13}$, 
A. W. Shaw$^{9,14}$, 
G. Sivakoff$^{9}$, 
A. J. Tetarenko$^{15}$, 
\newauthor F. Vincentelli$^{2}$ and 
R. Wijnands$^1$ \\
$^1$ Anton Pannekoek Institute for Astronomy, University of Amsterdam, Science Park 904, 1098 XH, Amsterdam, the Netherlands\\
$^2$ Physics \& Astronomy, University of Southampton, Southampton, Hampshire SO17 1BJ, UK\\
$^3$ Instituto de Astrof\'isica de Canarias, 38205 La Laguna, Tenerife, Spain\\
$^4$ Departamento de Astrof\'\i{}sica, Universidad de La Laguna, E-38206 La Laguna, Tenerife, Spain\\
$^5$ Department of Physics and Astronomy, Michigan State University, East Lansing, MI 48824, USA\\
$^6$ International Centre for Radio Astronomy Research, Curtin University, GPO Box U1987, Perth, WA 6845, Australia\\
$^7$ Facultad de Ciencias Astron\'omicas y Geof\'{\i}sicas, Universidad Nacional de La Plata, Paseo del Bosque s/n, 1900 La Plata, Argentina \\
$^8$ Instituto Argentino de Radioastronom\'{\i}a (CCT-La Plata, CONICET; CICPBA), C.C. No. 5, 1894 Villa Elisa, Argentina\\
$^9$ Department of Physics, CCIS 4-183, University of Alberta, Edmonton, AB, T6G 2E1, Canada\\
$^{10}$ Department of Physics, Box 41051, Science Building, Texas Tech University, Lubbock, TX 79409-1051, USA\\
$^{11}$ Department of Physics and Astronomy, Wheaton College, Norton, MA 02766, USA\\
$^{12}$ Department of Astronomy \& Astrophysics, Atat\"{u}rk University, Erzurum, Turkey\\
$^{13}$ Center for Astro, Particle and Planetary Physics, New York University Abu Dhabi, PO Box 129188, Abu Dhabi, UAE\\
$^{14}$ Department of Physics, University of Nevada, Reno, 1664 N. Virginia Street
Reno, NV 89557, USA \\
$^{15}$ East Asian Observatory, 660 N. A'oh$\overline{o}$k$\overline{u}$ Place, University Park, Hilo, Hawaii 96720, USA \\
}

\begin{document}

\date{Accepted XXX. Received YYY; in original form ZZZ}

\pagerange{\pageref{firstpage}--\pageref{lastpage}} \pubyear{2019}

\maketitle

\label{firstpage}

\begin{abstract}
Swift J1858.6-0814 is a transient neutron star X-ray binary discovered in October 2018. Multi-wavelength follow-up observations across the electromagnetic spectrum revealed many interesting properties, such as erratic flaring on minute timescales and evidence for wind outflows at both X-ray and optical wavelengths, strong and variable local absorption, and an anomalously hard X-ray spectrum. Here, we report on a detailed radio observing campaign consisting of one observation at 5.5/9 GHz with the Australia Telescope Compact Array, and nine observations at 4.5/7.5 GHz with the Karl G. Jansky Very Large Array. A radio counterpart with a flat to inverted radio spectrum is detected in all observations, consistent with a compact jet being launched from the system. Swift J1858.6-0814 is highly variable at radio wavelengths in most observations, showing significant variability when imaged on 3-to-5-minute timescales and changing up to factors of 8 within 20 minutes. The periods of brightest radio emission are not associated with steep radio spectra, implying they do not originate from the launching of discrete ejecta. We find that the radio variability is similarly unlikely to have a geometric origin, be due to scintillation, or be causally related to the observed X-ray flaring. Instead, we find that it is consistent with being driven by variations in the accretion flow propagating down the compact jet. We compare the radio properties of Swift J1858.6-0814 with those of Eddington-limited X-ray binaries with similar X-ray and optical characteristics, but fail to find a match in radio variability, spectrum, and luminosity.
\end{abstract}



\section{Introduction}
\label{sec:introduction}

X-ray binaries, wherein a compact object accretes from an orbiting companion star, form excellent laboratories to study both the accretion and ejection of matter. Such systems are often classified based on the mass of the donor star, with low-mass X-ray binaries (LMXBs) typically having a donor star with a mass $< 1 M_{\odot}$. While the compact object in some LMXBs persistently accretes from its companion, the majority of systems are instead transients \citep[see for recent catalogues, e.g., ][]{corralsantana2016,tetarenko2016_watchdog}. These transient LMXBs mostly reside in a quiescent state, interspersed with accretion outbursts typically lasting weeks to months. During these outbursts, LMXBs can reach mass accretion rates around or exceeding the Eddington limit, increasing multiple orders of magnitude in X-ray luminosity over the course of days to weeks \citep[e.g.][]{done2007}.

X-ray binaries can also launch outflows from the accreted material, which can either take the form of collimated, relativistic jets, or slower, more massive, disk winds. In a typical black hole or neutron star LMXB outburst, the compact jet and disk winds are not observed simultaneously in radio and X-rays \citep[e.g.][]{fender2004, ponti2012_winds}. Instead, the compact radio jet is launched in the `hard' state, dominated by non-thermal, Comptonized X-ray emission \citep{corbel2000,dhawan2000,stirling2001}, while the winds are detected in X-rays \citep[if inclination allows;][]{ponti2012,higginbottom2017} in the `soft', disk-dominated state instead. In the soft state, the compact jet is quenched in black hole systems \citep{fender2004}. In neutron star systems, soft state jets are observed in some sources, while they are quenched in others \citep{millerjones2010, gusinskaia2016, diaztrigo2018}. The jet--wind dichotomy becomes less defined, however, when optical wavelengths are included: disk winds have been observed in the optical bands during the hard states of LMXBs, for instance in the black hole transients Swift J1357.2-0933 \citep{jimenez2019,charles2019} and MAXI J1820+070 \citep{munozdarias2019}. 

The relation between the wind and jet outflows is also more complex at the highest, super-Eddington accretion rates. At such rates, both outflow types are often seen simultaneously \citep[e.g.][]{neilsen2009,homan2016,munozdarias2016,motta2017,allen2018,vandeneijnden2019_chandra}. Super-Eddington accreting black holes often show flaring by several orders of magnitude on seconds to minutes time scales, in both X-rays and radio bands, where the radio variability is typically associated with the launch of discrete, expanding ejecta \citep[e.g.][]{rupen2003,rupen2004,fender09,tetarenko2017_v404, millerjones2019}. This origin can be inferred either indirectly from steep radio spectra, and a smoothing out of the variability with increasing delays towards lower observing frequencies, probing further down the outflow \citep{tetarenko2019}. Alternatively, it can be inferred more directly by resolving the ejecta with Very-long Baseline Interferometry \citep[e.g.][]{millerjones2019}. Similar behaviour is also seen in neutron star LMXBs when they accrete close to their Eddington limit \citep{fomalont2001,fomalont2001b,homan2016,motta2019}.

In October 2018, \citet{krimm2018} reported the discovery of a new Galactic X-ray transient with the \textit{Burst Alert Telescope} (BAT) aboard the \textit{Neil Gehrels Swift Observatory} (Gehrels et al. 2004; hereafter \textit{Swift}). This new source, named Swift J1858.6-0814 (hereafter Sw J1858), was quickly followed up at many wavelengths and also detected at UV \citep{kennea2018, bozzo2018}, optical \citep{vasilopoulos2018}, and radio \citep{bright2018} wavelengths. At X-ray energies, Sw J1858 showed an extremely hard spectrum up to 10 keV ($\Gamma < 1$) and a faint outburst, with fluxes up to $\sim 10^{-10}$ erg/s/cm$^2$ \citep[][]{reynolds2018}. On top of this relatively faint emission, Sw J1858 showed strong, brief X-ray flares \citep{ludlam2018,hare2019} where the observed flux could increase by more than an order of magnitude \citep[up to, for instance, $\sim 3.5\times10^{-9}$ erg/s/cm$^2$ in][]{hare2020}. Such flares which were also identified at optical wavelengths \citep{vasilopoulos2018,baglio2018,rajwade2018,rajwade2019,paice2018}. In addition, the X-ray spectrum showed evidence of strong local absorption \citep{reynolds2018,hare2020}, and outflows were observed in the optical spectrum \citep{munozdarias2020} and inferred from the X-ray spectrum as well (Buisson et al., \textit{submitted}).

Sw J1858 has been active for more than a year at the time of writing: it went into Sun constraint three weeks after the first report of the outburst, but appeared out of Sun constraint in a similar state three months later \citep{rajwade2019}. It has also remained active after a second Sun constraint period in 2019--2020, out of which it emerged in a softer, previously unobserved X-ray state in February 2020 \citep{buisson20a}. As noted by several authors \citep[i.e.][Buisson et al., \textit{submitted}]{ludlam2019,paice2018,reynolds2018,hare2020}, the rich phenomenology observed in the initial state of Sw J1858 overlaps with the properties of the transient black hole LMXBs V404 Cyg \citep{rodriguez2015} and V4641 Sgr \citep{wijnands2000}: the strong optical and X-ray flaring, X-ray spectral shape \citep[e.g.,][]{motta2017}, and evidence for optical winds is observed in all three sources \citep{munozdarias2016,munozdarias2018,munozdarias2020}. 

While Sw J1858 was in its new X-ray state since Feb 2020, \textit{NICER} observed several flares (one of which was also covered by \textit{NuSTAR}) that, upon closer inspection, turn out to be Type-I X-ray bursts \citep{buisson20b}. Such bursts are caused by runaway thermonuclear burning of accreted material on the surface of neutron stars \citep[see for reviews, e.g.,][]{lewin95,strohmayer06,galloway2017}. The bursts observed from Sw J1858 are identified as Type-I bursts from their profile, spectrum, and spectral evolution, and therefore unambiguously show that the accretor is a neutron star. One of the bursts reported by \citet{buisson20b} shows evidence for photospheric radius expansion, which implies a distance of $\sim 15$ kpc to Sw J1858 (assuming the burst reaches the Eddington luminosity; see also Buisson et al. \textit{in prep.}). We will adopt this distance for our work. 

Outside of its flares and before the second Sun constraint period, the outburst of Sw J1858 reached X-ray luminosities of $\lesssim 1.5\%$ (D/15 kpc)$^2$ $L_{\rm Edd}$ for a neutron star accretor. Given this low X-ray luminosity, we triggered a \textit{Swift} X-ray and Karl G. Jansky Very Large Array (VLA) radio monitoring program to study the outbursts of X-ray transients at low X-ray luminosity. This campaign was supplemented by an observation with the Australia Telescope Compare Array (ATCA). Here, we report the results of this study, focusing on the radio variability properties down to minute time scales. This paper presents the radio campaign in combination with \textit{Swift} monitoring, which acts to place the radio behaviour into the context of the full outburst.

\section{Observations and data analysis}
\label{sec:observations}

\subsection{Radio}
\label{sec:vla}

The radio observing campaign of Sw J1858 consisted of one initial observation with ATCA, followed by nine observations performed with the VLA. The ATCA observation was taken on November 8, 2018 between 04:58:09.9 and 11:08:59.9 UTC (project code C2601). The telescope was in its extended 6km (6B) configuration, with antenna 3 offline. We used PKS 1934-638 as the primary calibrator, while the nearby secondary calibrator for all observations was J1832-1035 ($\sim 7^{\rm o}$ separation). Measurements were recorded simultaneously at $5.5$ and $9$ GHz in standard continuum mode , with 2048 MHz bandwidth each. In this paper, we will refer to this observation as epoch 1. Details on all individual radio observations, including those discussed below, can be found in Table \ref{tab:obs}.

The nine VLA observations were part of two observing programs, SE0057 and 19A-495. The first six observations were performed in November 2018, before the first X-ray Sun constraint. The final three observations were obtained in February, March, and August 2019, respectively. The first eight observations lasted $1$ hour, including setup and calibration scans, and were taken in C-band using 8-bit mode, with two sub-bands centred at $4.5$ and $7.5$ GHz with $1024$ MHz bandwidth each. The final observation had the same setup but lasted for $3.5$ hours, maximizing the overlap with other observatories during a coordinated observing campaign. Depending on the start time of the observation, we used either J0137+3309 (3C 48) or J1331+3030 (3C 286) as the primary calibrator. We used the same secondary calibrator as for the ATCA observation. The observations were taken in different configurations, covering D, D$\rightarrow$C, C, C$\rightarrow$B, B, and A. Throughout this paper, we will refer to these nine observations as epochs 2 through 10. 

To calibrate and image the time-averaged radio observations, we followed standard procedures using the \textsc{Common Astronomy Software Application} \citep[\textsc{casa};][]{mcmullin2007} version 4.7.2. We used a combination of automatic flagging routines and careful manual inspection of the visibilities to remove RFI, and subsequently imaged Stokes I using the multi-scale multi-frequency \textsc{casa}-task \textsc{clean}. In order to maximize sensitivity while reducing the sidelobe-effects of a close-by background source, we use a Briggs weighting scheme with the robust parameter set to $0$. The source was significantly detected in each epoch in both bands, and we measured flux densities by fitting the point source with an elliptical Gaussian model with the size and shape of the restoring beam in the image plane using \textsc{imfit}. The 1$\sigma$ uncertainty on the flux density was determined as the rms over a close-by, source-free region. 

Given the strong X-ray and optical variability of Sw J1858, we also analysed the radio observations on shorter timescales to explore intra-observational variability. Due to their different instantaneous uv-plane coverage, we applied different approaches to the ATCA and VLA observation. Making use of its good instantaneous uv-plane coverage, we applied an image-plane approach to the VLA observations: using a custom \textsc{python}-script and manual checks of the output, we imaged the two observing bands ($4.5$ and $7.5$ GHz) on a $3$-minute time scale to measure the source's light curve. For ATCA, the East-West orientation makes image-plane analysis on time scales of minutes, at the flux density levels of Sw J1858, unfeasible. Therefore, we instead used uv-plane fitting with the \textsc{uvmultifit} task \citep{martividal2014}, using a $5$-minute time resolution. For both the VLA and ATCA, we attempted several time resolutions and inspected the results to find the time scale that best balances sensitivity with the ability to observe short-time-scale variations. We settled on a slightly lower ATCA time resolution, given the higher sensitivity of the VLA. The errors on the light curves were calculated in the same way as the averaged flux density in the image-plane analysis, and as the statistical fitting error in the uv-plane analysis. To assess to what degree any observed variability is intrinsic to the target, instead of atmospheric or instrumental effects, we also analysed a nearby background source on the same time scale for both the VLA and ATCA. 

Finally, for both the full and the time-resolved observations, we calculated the radio spectral index $\alpha$ (where $S_{\nu} \propto \nu^{\alpha}$). We estimated the error on $\alpha$ using a Monte-Carlo simulation, where we randomly draw flux densities at both observing frequencies and recalculate $\alpha$ in total $10^4$ times. The $1\sigma$ uncertainty on the spectral index was then calculated as the standard deviation of the resulting distribution of spectral indices. 


The VLA primary calibrator 3C 48, used in four epochs (see Table \ref{tab:obs}), has been undergoing flaring behaviour since January 2018\footnote{\href{https://science.nrao.edu/facilities/vla/docs/manuals/oss/performance/fdscale}{https://science.nrao.edu/facilities/vla/docs}\href{https://science.nrao.edu/facilities/vla/docs/manuals/oss/performance/fdscale}{/manuals/oss/performance/fdscale}}, which can affect the flux scale at the $\sim 5$\% level. To check whether this affects our observations, we inspected the output \textsc{casa} logs of the \textsc{fluxscale} task and compared the calibrated flux density of the secondary calibrator between the nine epochs. While the ATCA calibrator database\footnote{\href{https://www.narrabri.atnf.csiro.au/calibrators/}{https://www.narrabri.atnf.csiro.au/calibrators/}} shows that J1832-1035 remains at a relatively stable $5.5$ GHz flux density over time, we measure a slightly decreased flux density during epochs 3 and 4: $\sim 1.36$ Jy, compared to $\sim 1.45$--$1.50$ Jy in the remaining epochs, regardless of primary calibrator. Therefore, the measured average flux densities in those two epochs might be underestimated by $\sim 7$--$10$\%. However, this will not affect the time resolved analysis within observations that we present in this paper.

\subsection{X-ray}

\textit{Swift} performed an extensive X-ray monitoring campaign of Sw J1858 under ObsIDs $10955$, $10970$, and $88868$. Obervations were taken both in the Photon Counting (PC) and Windowed Timing (WT) modes. Before 22 November 2018, when the target went into Sun constraint, 13 X-ray Telescope (XRT) observations with exposures up to $2$ ks were performed. The monitoring then resumed on 19 February 2019, as the source was actively accreting as it appeared from Sun constraint, continuing until early August 2019. As we focus on the radio properties and variability of Sw J1858, we do not perform a detailed X-ray analysis of all \textit{Swift} observations, but instead focus on measuring the X-ray flux around the radio observations. For more detailed analysis of the X-ray spectrum of Sw J1858, we refer the reader to \citet{hare2020} and Buisson et al. (\textit{submitted}). 

Due to the faintness of the source, we do not use data from either \textit{Swift}-BAT or \textit{Monitor of All-sky X-ray Image (MAXI)} observations in this work. Instead, we use the available \textit{Swift}/XRT monitoring data to extract a long term X-ray light curve of the source, in order to place our radio observations in the context of the full outburst. In addition, we extract and fit spectra of the observations taken closest to our radio epochs, in order to measure the quasi-simultaneous X-ray flux. For every \textit{Swift} observation taken closest to a radio observation, we also consider a $100$-second resolution light curve to estimate the amount of X-ray variability at that time. All analysed X-ray data are listed in Table \ref{tab:obs}.

The only Swift/XRT observation within four days of the ATCA and first VLA observation (i.e. epochs 1 and 2) was merely $30$ seconds long. Therefore, for that observation we converted the X-ray count rate to a flux assuming a X-ray spectral shape typical of Sw J1858 in the remainder of the outburst (see Section \ref{sec:xray_fluxes}). For epochs 3 to 7, all performed before the Sun constraint when \textit{Swift} observed nearly every day, the analysed X-ray spectra where taken within $24$ hours of the radio observation. For epochs 8 \& 9, performed after the first Sun constraint, \textit{Swift} monitoring was more sparse. Therefore, the radio and X-ray observations at these later times are typically separated by $\sim 1.5$--$2$ days. Finally, three consecutive \textit{Swift} observations were taken simultaneously with the tenth and final epoch.  

\textit{Swift}/XRT spectra, with corresponding background spectra and response files, and light curves (both long-term and of single observations) were obtained using the online data products generator\footnote{\href{http://www.swift.ac.uk/user\_objects/}{http://www.swift.ac.uk/user\_objects/}} (Evans et al. 2007, 2009). We fit the X-ray spectra with \textsc{xspec} v.12.10.1, setting the abundances to \citet{wilms2000} and cross-sections to \citet{verner1996}. We account for interstellar absorption using the \textsc{tbabs} model. Given the low count rates and short observations, we used W-statistics as our fit statistic \citep{cash1979}. Fluxes with errors were calculated using the convolution model \textsc{cflux}. All reported errors are quoted at the $1\sigma$ level, unless stated otherwise.

\section{Results}
\label{sec:results}

\begin{table*}
	\centering
	\caption{Overview of the ten radio observations of Sw J1858, with their corresponding quasi-simultaneous \textit{Swift}/XRT X-ray observations. For epoch 1 through 8 and 10, the time difference is less than one day. For epoch 9, the two surrounding X-ray epochs are both separated by a $\sim 1.3$ day difference. Therefore, we consider both for our X-ray flux measurements. For epoch 10, the three listed \textit{Swift} observations were back-to-back. All listed MJDs represent the start of the observations. *In these two epochs, flaring in 3C 48 might have lead to an underestimate of the target flux density by $\sim 7$--$10$\%, see Section \ref{sec:vla} for details.}
	\label{tab:obs}
	\begin{tabular}{lllllllll} %
Epoch & Observatory & Radio MJD & Primary calibrator & Config. & \textit{Swift} ObsID & XRT mode & \textit{Swift} MJD & XRT Exposure [s] \\ 
\hline \hline
1 & ATCA & 58430.207 & PKS 1934-638 & 6B & \multirow{2}{*}{00010955003} & \multirow{2}{*}{WT} & \multirow{2}{*}{58432.619} & \multirow{2}{*}{30}\\
2 & VLA & 58431.954 & J0137+3309 (3C 48) & D & & & &\\ \hline
3 & VLA & 58436.072 & J0137+3309 (3C 48)* & D$\rightarrow$C & 00010955004 & WT & 58436.478 & 1870\\
4 & VLA & 58437.060 & J0137+3309 (3C 48)* & D$\rightarrow$C & 00010970002 & PC & 58437.465 & 2035\\
5 & VLA & 58438.835 & J1331+3030 (3C 286) & D$\rightarrow$C & 00010970003 & PC & 58439.390 & 1630\\
6 & VLA & 58443.748 & J1331+3030 (3C 286) & C & 00010955008 & PC & 58444.304 & 1960\\
7 & VLA & 58443.922 & J0137+3309 (3C 48) & C & 00010955008 & PC & 58444.304 & 1960\\
8 & VLA & 58534.498 & J1331+3030 (3C 286) & C$\rightarrow$B & 00010970008 & PC & 58537.342 & 995\\
9 & VLA & 58566.479 & J1331+3030 (3C 286) & B & 00010970023/24 & PC & 58565.104/58567.754 & 995/945\\
10 & VLA & 58701.035 & J1331+3030 (3C 286) & A & 00010970043/44/45 & PC & 58701.032 & 2200  \\
\hline
	\end{tabular}
\end{table*}

\begin{table*}
	\centering
	\caption{Radio and X-ray flux (densities) and spectral information. For each radio epoch, we list the low band ($4.5$ or $5.5$ GHz for VLA or ATCA, respectively) and high band ($7.5$ or $9$ GHz) flux density $S_{\nu}$ and the radio spectral index $\alpha$, where $S_{\nu} \propto \nu^{\alpha}$. In the X-rays, we list the measured 0.5 -- 10 keV flux, spectral index $\Gamma$, and W-statistic with degrees of freedom. Each Swift/XRT spectrum was fitted with a \textsc{tbabs*powerlaw} model, but $N_H$ was fixed to zero as no absorption was required for any of the fits. *In these two epochs, flaring in 3C 48 might lead to an underestimate of the target flux density by $\sim 7$--$10$\%, see Section \ref{sec:vla} for details.}
	\label{tab:mean_results}
	\begin{tabular}{lcccccc} %
\multirow{2}{*}{Epoch} & \multicolumn{2}{c}{Radio flux density [$\mu$Jy]} & \multirow{2}{*}{Radio spectral index} & $0.5$--$10$ keV X-ray flux & \multirow{2}{*}{$\Gamma$} & \multirow{2}{*}{$W_{\rm stat}$ (dof)}\\ 
& Low band & High band & & [erg/s/cm$^2$] & & \\
\hline \hline
1 & $422 \pm 15$ & $505 \pm 15$ & $0.36 \pm 0.11$ & \multirow{2}{*}{$ (2.9 \pm 0.4_{\rm syst} \pm 1.0_{\rm pois} )\times 10^{-11} $} & \multicolumn{2}{c}{\multirow{2}{*}{N/A (see Section \ref{sec:xray_fluxes})}} \\ 
2 & $499 \pm 12 $ & $607 \pm 9 $ & $0.40 \pm 0.18 $ & &  & \\
3* & $233 \pm 10 $ & $265 \pm 10 $ &$0.25 \pm 0.16 $ & $ (1.0 \pm 0.1)\times10^{-11} $ & $1.0\pm 0.2$ & 250.6 (265) \\
4* & $279 \pm 10 $ & $285 \pm 12 $ &$0.05 \pm 0.14 $ & $ (9.5 \pm 1.5)\times10^{-12} $ & $-0.1\pm 0.2$ & 127.5 (106) \\
5 & $313 \pm 9 $ & $411 \pm 8 $ &$0.55 \pm 0.24 $ & $ (2.3 \pm 0.2)\times10^{-11} $ & $0.2\pm 0.1$ & 181.8 (221) \\
6 & $210 \pm 12 $ & $275 \pm 11 $ &$0.55 \pm 0.28 $ & $ (1.7 \pm 0.1)\times10^{-10} $ & $0.45\pm 0.05$ & 451.3 (507) \\
7 & $172 \pm 9 $ & $233 \pm 9 $ &$0.62 \pm 0.30 $ & $ (1.7 \pm 0.1)\times10^{-10} $ & $0.45\pm 0.05$ & 451.3 (507) \\
8 & $187 \pm 12 $ & $184 \pm 9 $ &$0.00 \pm 0.19 $ & $ (2.1 \pm 0.3)\times10^{-11} $ & $0.0\pm 0.2$ & 112.7 (108)\\
9 & $130 \pm 10 $ & $182 \pm 8 $ & $0.66 \pm 0.32 $ & $ (1.2 \pm 0.2)\times10^{-11} $ & $0.5\pm 0.2$ & 98.1 (91) \\
10 & $166 \pm 9$ & $176 \pm 6$ & $0.11 \pm 0.13$ & $(3.1 \pm 0.2) \times 10^{-11}$ & $0.2 \pm 0.1$ & 256.4 (285) \\
\hline
	\end{tabular}
\end{table*}

\subsection{Radio detection and position}

\begin{figure}
	\includegraphics[width=\columnwidth]{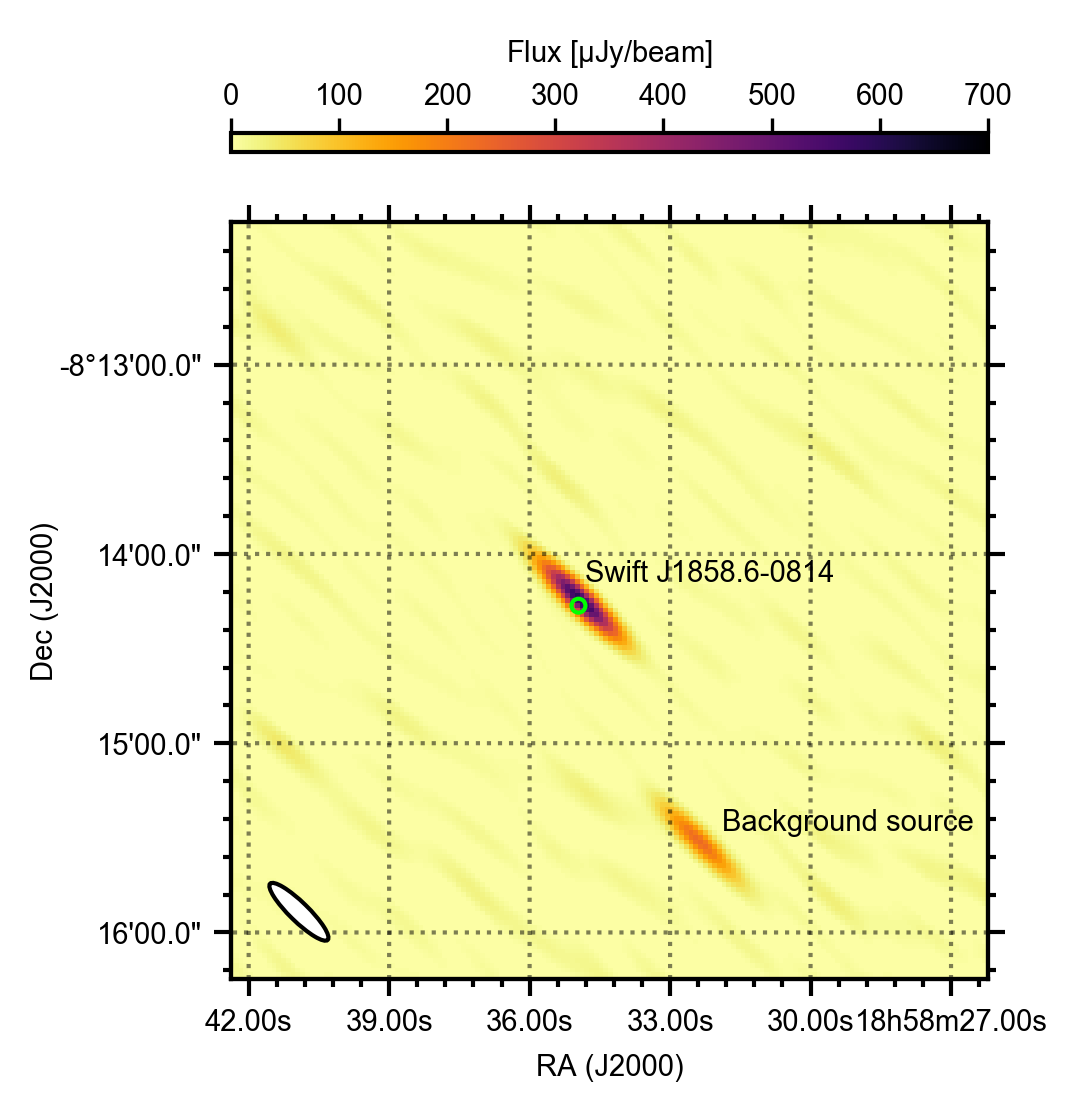}
    \caption{Detection image of epoch 2 at $7.5$ GHz. Note the nearby flat-spectrum background source, used to assess the systematic short-time-scale variability of the entire field. The green circle shows the $90$\% confidence region on the \textit{Swift} X-ray position.}
    \label{fig:radiodetection}
\end{figure}

Sw J1858 is significantly detected at both radio observing frequencies in every observation. In Figure \ref{fig:radiodetection}, we show the radio detection of Sw J1858 during the second epoch at $7.5$ GHz (observed with the VLA), when the source was radio brightest. The image shows a radio counterpart consistent with the \textit{Swift}/XRT source. A second, nearby radio source is also visible at a separation of $\sim 2'$, which might contribute to the structured noise reported by AMI-LA \citep{bright2018}. The best fit radio position was measured from the tenth, and final, epoch (7.5 GHz):\\

RA $=$ $18$h $58$m $34.9100$s $\pm$ $0.0027$s 

Dec $=$ $-08^{\rm o}$ $14$' $14.958$" $\pm$ $0.037"$\\

\noindent where the error is inferred from the typical astrometric accuracy of the VLA ($10$\% of the synthesised beam). This position is fully consistent with the Swift/XRT position reported by \citet{kennea2018}. We used the final epoch for this calculation, since the extended A-configuration during this observation yields the smallest statistical uncertainty. While the target was relatively faint during this observation, the high signal-to-noise ($S/N >> 10$) still ensured the positional accuracy was limited by $10\%$ of the beam size. 

\subsection{Long term light curves}

\begin{figure*}
	\includegraphics[width=0.95\textwidth]{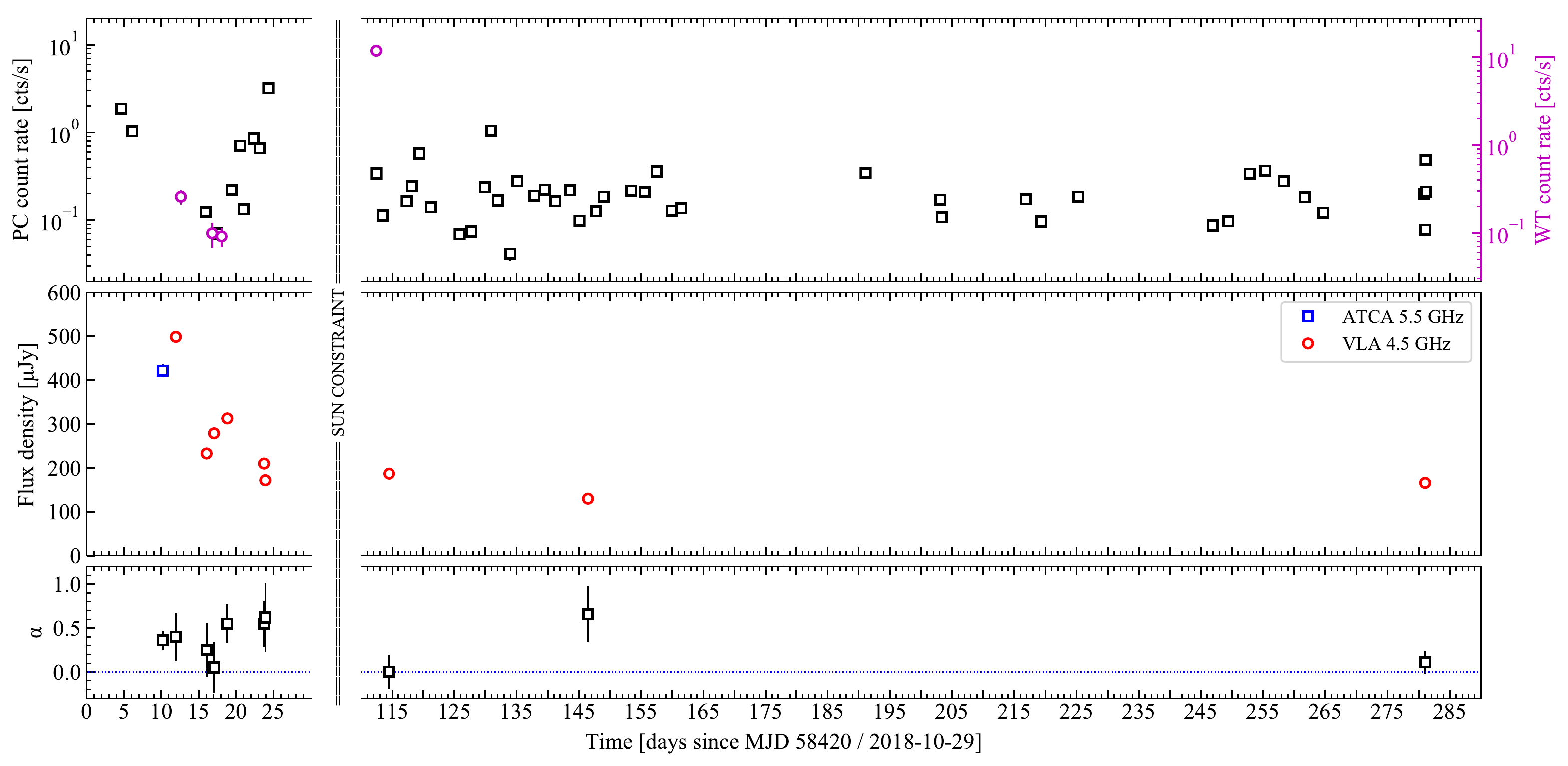}
    \caption{The long term X-ray, radio, and radio spectral index light curves of the 2018/2019 outburst of Sw J1858. Roughly 30 days after the source's discovery, Sw J1858 went into X-ray Sun constraint for $\sim$ 80 days. The top panel shows the \textit{Swift}/XRT count rate in PC mode (black squares) and WT mode (purple circles). The PC and WT axes are offset by $39\%$, which represent the difference in count rate for the same flux (measured from observations 00010955004 and 00010970002; see Tables \ref{tab:obs} and \ref{tab:mean_results}). Note that in PC mode, the markers are larger than the uncertainties. The middle panel shows the radio flux density measured with the VLA ($4.5$ GHz; red circles) and ATCA ($5.5$ GHz; blue squares). Finally, the bottom panel shows the calculated radio spectral index $\alpha$. While clear variations in X-ray count rate are visible, no clear rise-decay outburst shape can be distinguished. The radio light curve does show a gradual decrease over time.}
    \label{fig:longtermlc}
\end{figure*}

\begin{figure*}
	\includegraphics[width=\textwidth]{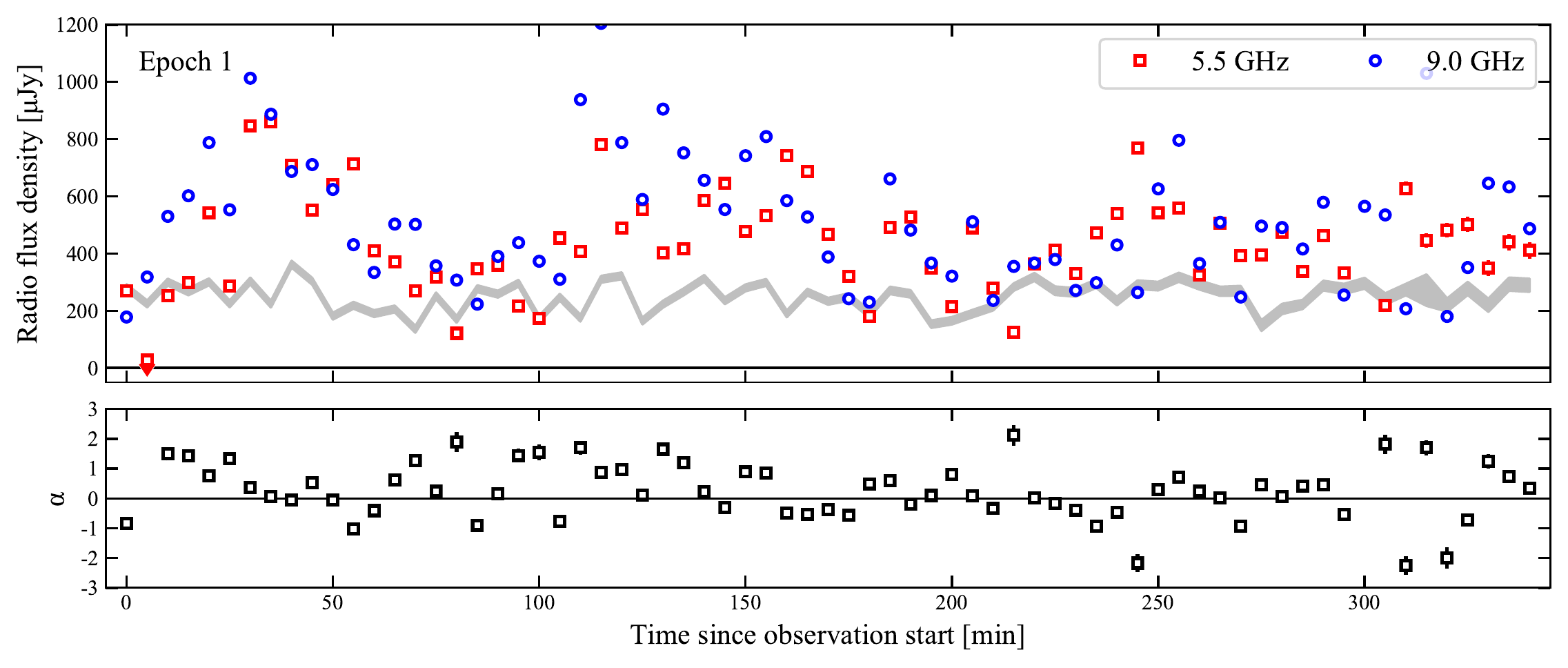}
    \caption{Light curves of the $5.5$ (red squares) and $9.0$ (blue circles) GHz flux density and the radio spectral index $\alpha$ at a five-minutes resolution, for the ATCA observation (epoch 1). The grey band shows the $9$-GHz light curve of the nearby flat-spectrum background source.}
    \label{fig:ATCAlcs}
\end{figure*}

\begin{figure*}
	\includegraphics[width=\textwidth]{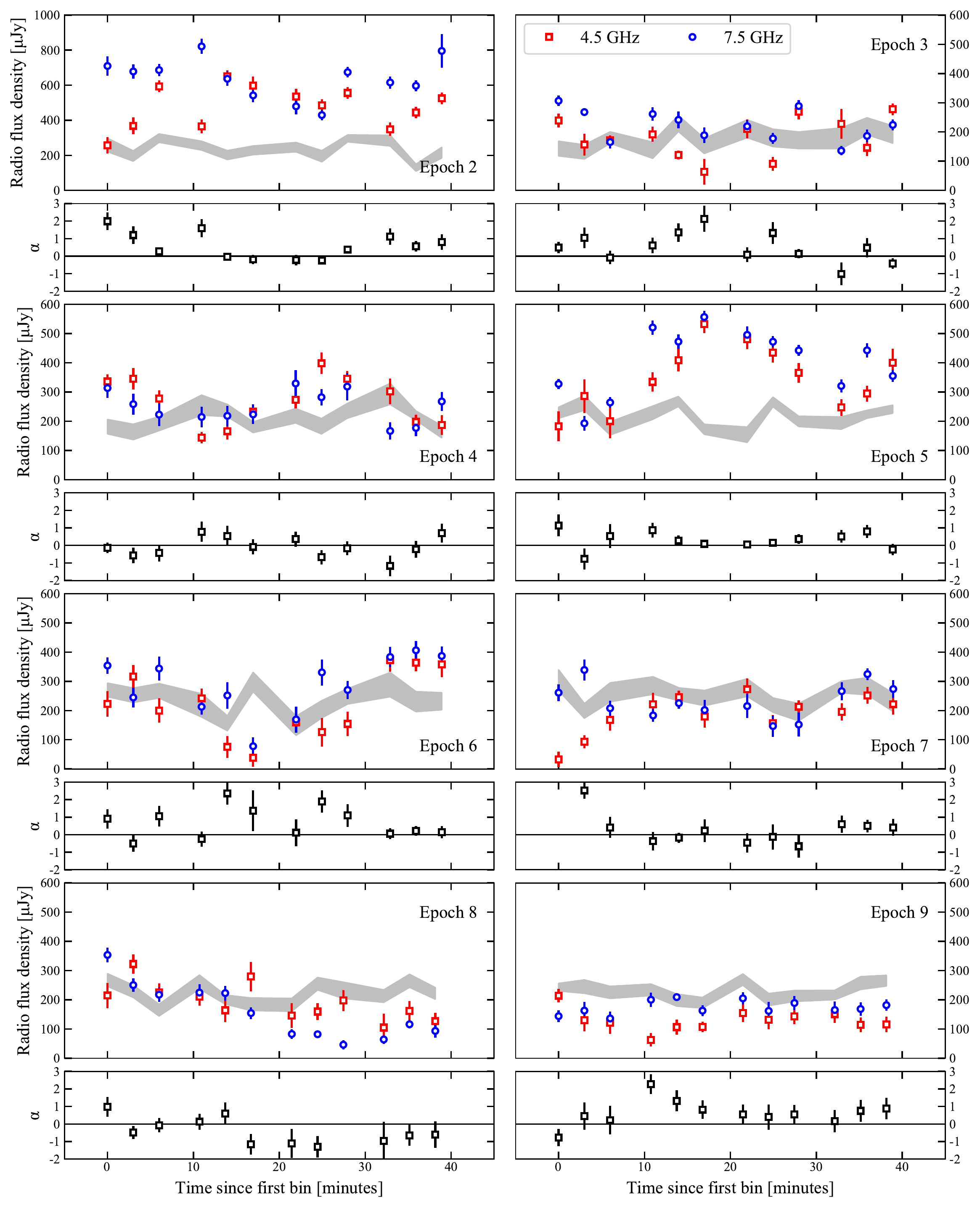}
    \caption{Light curves of the $4.5$ (red squares) and $7.5$ (blue circles) GHz flux density and the radio spectral index $\alpha$ at a three minute time resolution, for the first eight VLA observations (epoch 2 -- 9). Note the different vertical range in the upper left panel. The spectral index panels correspond to the epoch shown above it. The grey bands show the 1$\sigma$ band for the background source visible in Figure \ref{fig:radiodetection}.}
    \label{fig:obslcs}
\end{figure*}

\begin{figure*}
	\includegraphics[width=\textwidth]{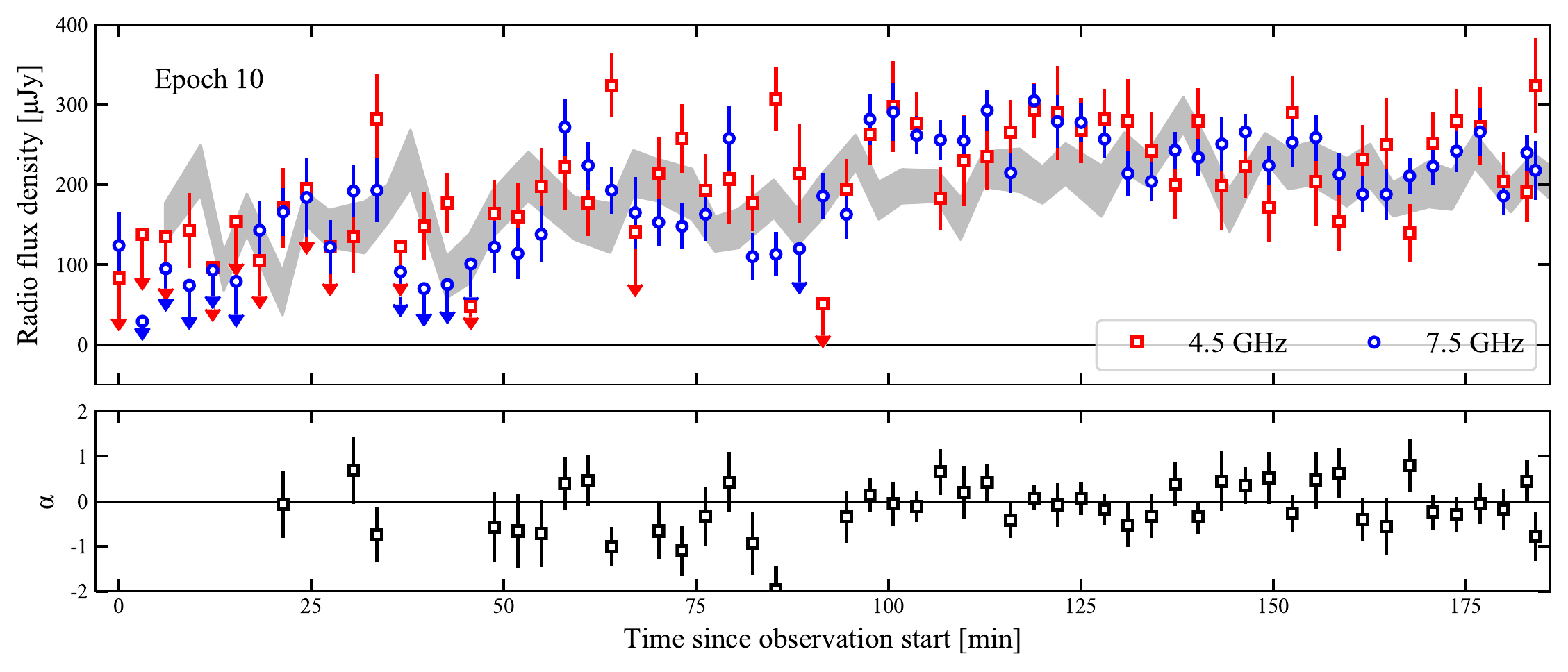}
    \caption{Same as Figure \ref{fig:obslcs}, for epoch 10. Note the difference in observation length with respect to the other VLA epochs. At the start of the observation, the source is not significantly detected in every $3$-minute time bin; therefore, we only calculate the spectral index $\alpha$ for time bins where the source is detected at both $4.5$ and $7.5$ GHz.}
    \label{fig:HSTlcs}
\end{figure*}

\begin{figure}
	\includegraphics[width=\columnwidth]{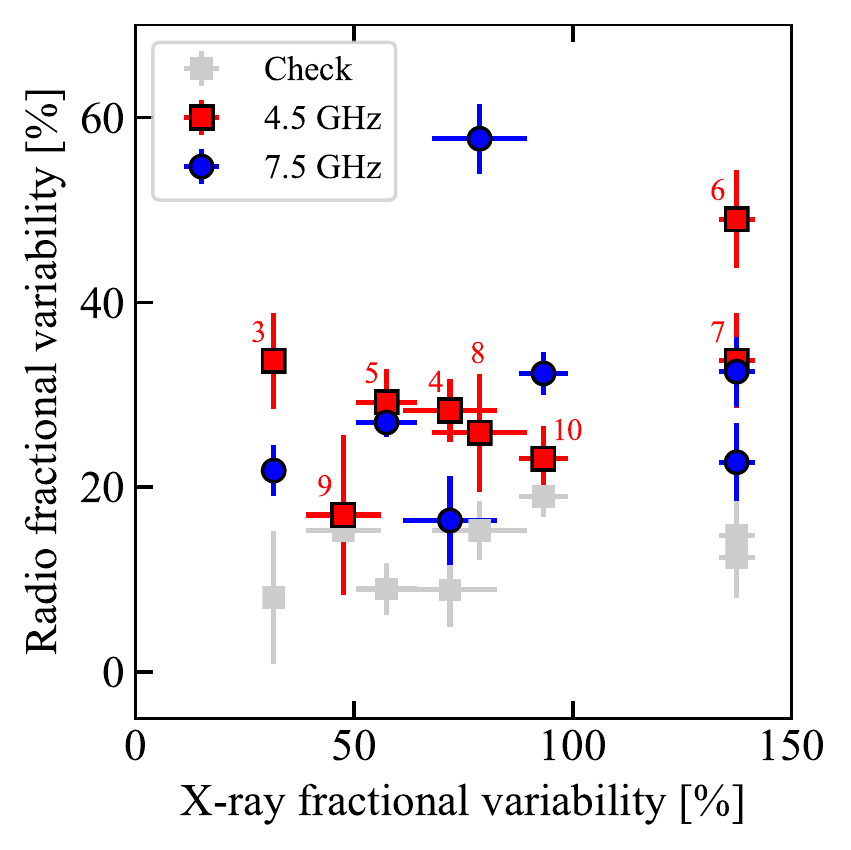}
    \caption{A comparison of the fractional variability in radio and X-ray at a $3$-minute and $100$-second time resolution, respectively. The red numbers correspond to the epoch number. The grey points indicate the background source, showing the typical level of variability that is not intrinsic to Sw J1858. Note that we plot these grey points as a function of X-ray fractional variability of Sw J1858, as no X-ray observations of the background source were taken.}
    \label{fig:Fvarcomp}
\end{figure}

\begin{figure*}
	\includegraphics[width=\textwidth]{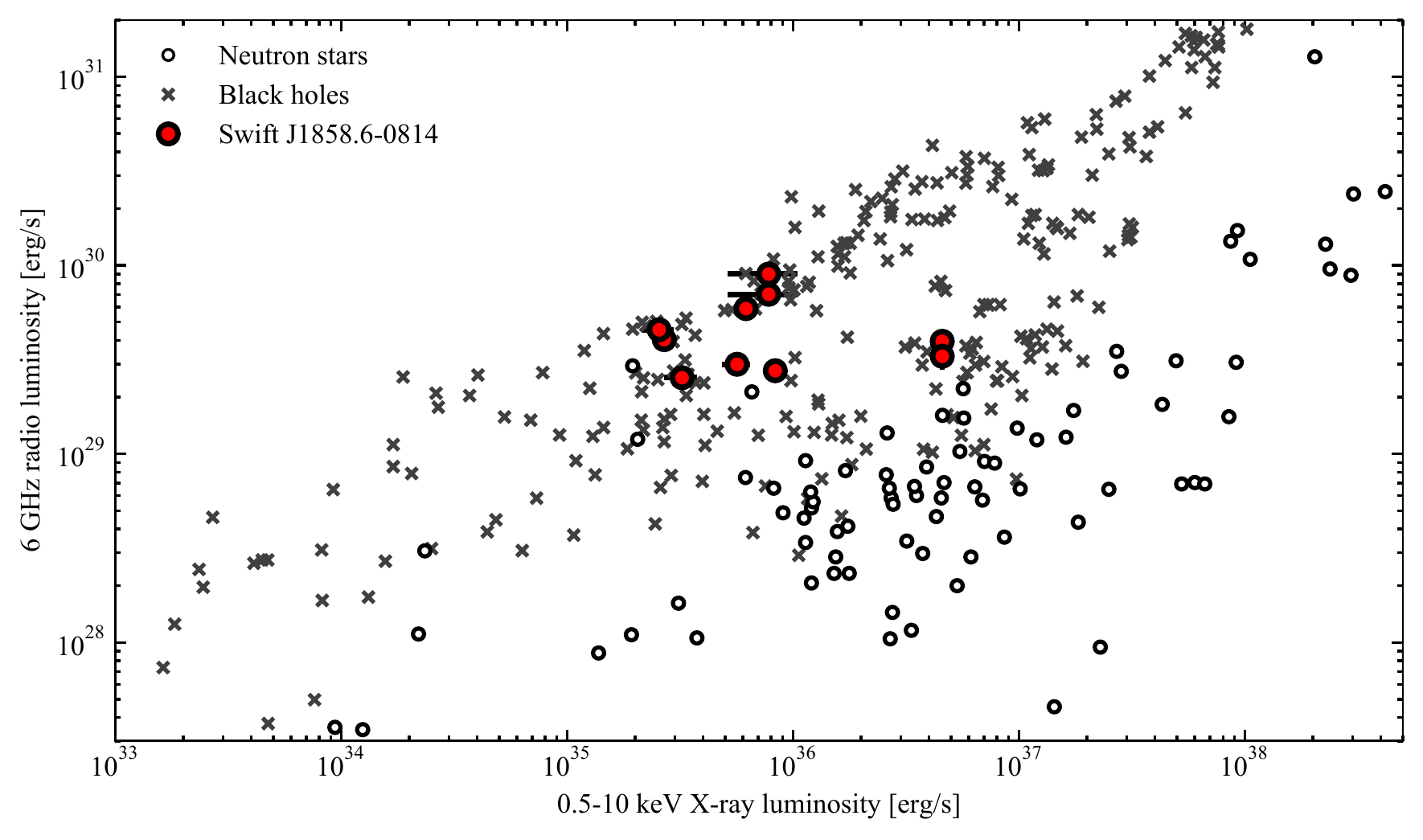}
    \caption{The X-ray -- radio luminosity plane for black hole and neutron star X-ray binaries. The grey crosses show black hole (candidates), while the circles show different classes of accreting neutron stars. We show the average $0.5-10$ keV X-ray and inferred $6$-GHz radio luminosities of Sw J1858 for a $15$ kpc distance. The two brightest radio points are epochs 1 and 2, which share the same X-ray observation. Note that the errors do not reflect the variability within observations, nor account for the non-simultaneity of the X-ray and radio observations and the variability between \textit{Swift} observations (Sec. \ref{sec:results_lxlr}). The comparison sample was taken from \citet{gallo2018}.}
    \label{fig:lxlr}
\end{figure*}

\begin{figure}
	\includegraphics[width=\columnwidth]{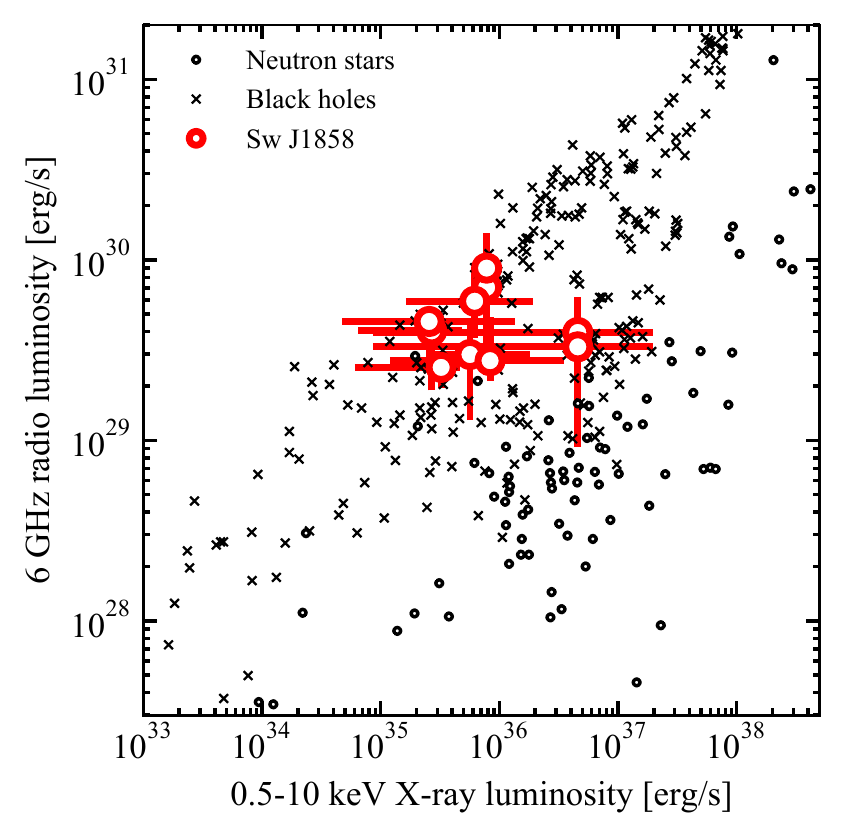}
    \caption{The X-ray -- radio luminosity plane with Sw J1858 shown in red circles. Here, the error bars reflect the variability range in X-ray count rate (on $100$ sec time scales) and radio flux density (on $3$ min time scales) during each observation.}
    \label{fig:lxlr_var}
\end{figure}

In Figure \ref{fig:longtermlc}, we show the long-term X-ray and radio light curves, and the radio spectral index as a function of time. Between $30$ and $110$ days after the outburst detection by BAT, Sw J1858 was in Sun constraint for \textit{Swift}. The X-ray count rate light curve (top) reveals strong variations between observations, changing more than two orders of magnitude. However, no clear outburst profile, with a distinct rise and decay, can be seen. The apparent variability between observations is affected by the presence of X-ray flaring in Sw J1858, which, in combination with the short Swift exposures, boosts the count rate in observations covering one or more flares. After day $\sim 135$, the dynamic range between observations decreases, with all count rates lying within one dex. At the time of writing, Sw J1858 still remains active, much longer than for instance the 2015 outburst of V404 Cyg that it was compared to early on in its outburst \citep{ludlam2018,paice2018,reynolds2018}

The radio light curve, shown in the middle panel, does show an initial peak that decays over time. However, again, the source does not monotonically decay in radio flux density. The radio sampling becomes sparse after the Sun constraint period, but Sw J1858 never decays below $\sim 170$ $\mu$Jy in our observations. All plotted radio flux densities, at both observing frequencies, are listed in Table \ref{tab:mean_results} as well. Finally, we show the radio spectral index as a function of time. In all observations, the spectral index is either consistent with zero or positive, implying a flat or inverted spectrum, respectively. Such a radio spectral shape in an X-ray binary system is consistent with the presence of a compact, steady jet \citep[e.g.][]{blandford79, corbel2004, fender2004, kalemci2005, russell14}. There appears to be no systematic evolution of the spectral index as the radio brightness decays.

\subsection{X-ray flux measurements}
\label{sec:xray_fluxes}

Alongside the radio flux densities and spectral shape, we also list the measured X-ray fluxes and spectral fit parameters in Table \ref{tab:mean_results}. To measure the \textit{observed} fluxes, we fitted every spectrum with a simple phenomenological power law model, \textsc{tbabs*po}. First fitting all spectra jointly to measure the absorption column $N_H$, we find that -- \textit{when using this model} -- the data does not require the inclusion of interstellar absorption. Hence, we fix $N_H$ to zero in the fits to individual spectra. We stress that this value does not imply a zero column density towards the target; instead, it arises from our non-physical model. Indeed, the measured power law indices are very hard ($-0.1 < \Gamma < 1.0$), also as a result of our model. 

Considering more complex spectral models, Reynolds et al. (2018) showed how \textit{Swift} observation 00010955004 (used in epoch 3) can be better fitted with a complex local absorption model, similar to that seen in V404 Cyg and V4641 Sgr \citep{motta2017,morningstar2014}. Our simple power law approach, however, can also be applied to the lower-flux spectra; for those spectra, we find that the local absorption model returns large degeneraries between parameters as the spectra are overfitted. Moreover, as we discuss in Section \ref{sec:discuss_xrays}, there is no systematic difference between the flux measurements obtained from the two models in the observations where both can be applied. Since we aim to determine the observed flux, and discuss the effects of local absorption mostly qualitatively in Section \ref{sec:eddington}, we therefore decided on using the power law model in the analysis of all X-ray spectra. 

Two sets of observations, namely radio epochs 1 and 2 (which are the closest to first \textit{Swift} observation), and epoch 9, stand out. Firstly, for epochs 1 and 2, the closest Swift/XRT spectrum is merely 30 seconds long. Therefore, instead of fitting the spectrum, we converted the measured WT-mode count rate to a flux. For this purpose, we measured the power law index in the preceding and following X-ray observation ($\Gamma = 0.06 \pm 0.15$ and $\Gamma = 0.19 \pm 0.15$, respectively). We then selected the most extreme values within their $1-\sigma$ error ranges and used \textsc{webpimms} to convert the count rate to a minimum and maximum flux. We select the average of those as the measured flux and refer to the range as the systematic error in Table \ref{tab:mean_results}. Finally, we also include the Poisson error on the count rate. Secondly, the ninth radio epoch was surrounded by two Swift/XRT observations with approximately the same time difference of $\sim 1.3$ days. Therefore, we fitted both spectra and measured fluxes for both observations. We measure consistent fluxes in these two observations; in the remainder of this work, we adopt the value of the X-ray observation taken after the radio epoch (ObsID 0001097024, ignoring ObsID 0001097023).

\subsection{Short-time-scale variability}
\label{sec:intravar}

Given the flaring behaviour of Sw J1858 in both X-ray and optical bands, we set out to investigate whether similar extreme variability is present at radio frequencies. Therefore, we imaged the radio counterpart at both observing frequencies on a 5/3 minute (ATCA/VLA) time scale. Time resolving at this time scale allowed us to study quick changes in both flux and spectral index, without the increased uncertainties masking any variability. In Figures \ref{fig:ATCAlcs}, \ref{fig:obslcs}, and \ref{fig:HSTlcs}, we show the resulting light curves for each epoch. We then calculated the spectral index $\alpha$ in each time bin, shown in Figures \ref{fig:ATCAlcs}, \ref{fig:obslcs}, and \ref{fig:HSTlcs} as well. The VLA light curves also show, in grey, the $1\sigma$ bands of the flux density of the nearby background source. These bands show to what extend the image-plane analysis is affected by atmospheric or instrumental effects and vary typically on or slightly above the level of the target flux densities' uncertainties. 

The ATCA light curve in Figure \ref{fig:ATCAlcs} shows a highly variable target at both observing frequencies during the $\sim 6$ hour observing time. At the start of the observation, a flare is observed, rising and decaying on a time scale of tens of minutes. During the rise and decay of this flare, the spectral index gradually decays from inverted ($\alpha > 1$) to flat ($\alpha \approx 0$). The flux densities across the observation span a dynamical range of more than four, varying between less than $200$ $\mu$Jy to more than $800$ $\mu$Jy on time scale of tens of minutes. We stress that we are plotting the statistical error on the flux density only, as determined from the uv-plane fit to the point source, which is of the same order as the image-plane RMS in the averaged observation (i.e. $15$ $\mu$Jy). Therefore, the uncertainties are often smaller than the markers. We also show the light curve of the background source, which gives an indication of the systematic uncertainties on the flux density measurements (assuming it is intrinsically constant). 

Turning to the VLA light curves in Figure \ref{fig:obslcs}, we observe significant intrinsic variability compared to the background source at either one or both radio frequencies in most epochs. For instance, epoch 4 shows a sequence of decay, rise and decay within the single $40$-minute observation. Alternatively, we observe a clear radio flare in epoch 5, rising from $\sim 200$ to $\sim 600$ $\mu$Jy in less than 15 minutes, than decaying over approximately 20 minutes.  Epoch 8 shows a strong decay at 7.5 GHz, from $\sim 350$ $\mu$Jy down to $\sim 50$ $\mu$Jy in less than $30$ minutes. The variability in the other epochs appears more erratic, of a seemingly stochastic nature. We do not observe any strong flares on the time scale of $\sim 3$ minutes, i.e. in a single time bin. Due to the image-plane approach for these VLA observations, the uncertainties on the individual measurements are significantly larger than for the ATCA epoch. 

With its long total observing time, the final VLA observation (epoch 10; Figure \ref{fig:HSTlcs}) offers a complementary view to epochs 2 to 9: during this observation, the source was not significantly detected in every $3$-minute time bin. Particularly around the start of the observation, Sw J1858 hovered around flux densities three times the typical short-time-scale RMS sensitivity. Therefore, we only calculate the spectral index in time bins where the source is significantly detected at both $4.5$ and $7.5$ GHz. During the second half of the observation, Sw J1858 reaches a steady level above the detection threshold, with a relatively flat spectral index. At that point, it does however not appear significantly more variable than the background source. These results show that the behaviour of Sw J1858 in the 40 minute observations during epochs 2--9 is not necessarily stationary and can change quite rapidly. 

To assess the level and significance of the radio variability, we calculated the fractional variability $F_{\rm var}$ of the light curves of Sw J1858 at both frequencies, and of the background source, following the description in \citet{vaughan2003}. The fractional variability of a light curve $x(t)$ is defined as $F_{\rm var} = \sqrt{(S^2 - \overline{\sigma^2_{\rm err}})/\overline{x}^2}$, where $S^2$ is the light curve variance and $\overline{\sigma^2_{\rm err}}$ is the mean square error on the count rate. The calculated values of $F_{\rm var}$ are listed in Table \ref{tab:variability}. For all epochs but 1, 8 and 10, we observe stronger variability at the lower frequency. In most VLA epochs, the observed fractional variability also exceeds that of the background source, indicating that Sw J1858 was intrinsically variable. In epoch 3 and 5--8, this is the case at both frequencies, while for epochs 4, 9, and 10, only at a single frequency $F_{\rm var}$ significantly exceeds the background source. Only in epoch 2, Sw J1858 and the background source show consistent levels of variability. Note that for epoch 9, $F_{\rm var}$ could not be calculated at $7.5$ GHz as the observed variations in flux density where smaller than the average error on the data \citep[$S^2<\overline{\sigma^2_{\rm err}}$ in the notation of][]{vaughan2003}. Finally, the much lower uncertainties on the $F_{\rm var}$ measurements for epoch 1 are caused by using only the statistical uncertainty on the uv-plane fit.

In Figures \ref{fig:ATCAlcs}, \ref{fig:obslcs}, and \ref{fig:HSTlcs}, we also show the radio spectral index as a function of time. We do not observe any optically thin, i.e. $\alpha \approx -0.7$, radio flares, that are associated with the launch of individual ejecta. For instance, in the flare observed at the start of epoch 1, the spectrum is flat to inverted, while during the flare observed in epoch 5, the spectrum is tightly constrained to be flat. With the exception of the decreasing spectral index during the flare in epoch 1, we generally observe little structured evolution in the spectral shape as a function of time, as might for instance result from lags between the two observing bands. Similarly, we do not find evidence for a relation between the radio flux and spectral shape. Finally, in several epochs, most notably 2, 7, and 8, we observe strongly inverted radio spectra ($\alpha > 2$), turning more flat as the $4.5$ GHz flux rises towards the $7.5$ GHz flux.  

We search for a time delay between the $4.5$ and $7.5$ GHz light curves
using the method described in Section 3.4 of \citet{buisson2017}. To remove the long-term delay during the outburst, we subtracted the mean flux per observation and frequency band. The measured time lag between the full campaign VLA light curves is $\tau = -110 \pm 360$ seconds, where we quote the $1$-sigma uncertainty and positive values imply that the $7.5$ GHz band lags behind the $4.5$ GHz band. Hence, the observations are consistent with no time lag between the two frequencies. 

Finally, since strong variability is also observed in X-rays, we investigated whether the levels of X-ray and radio variability in Sw J1858 could be linked. For this purpose, we extracted $100$-second time resolution $0.5$--$10$ keV light curves of all \textit{Swift} observations listed in Table \ref{tab:obs}, except for the 30-second observation 00010955003 -- which means that we do not include epoch 1 and 2 in this comparison. We then calculated the fractional X-ray variability, which we compared with the radio variability in Figure \ref{fig:Fvarcomp}. From this comparison we can see that no clear relation exists between the fractional variability in X-ray and at either radio observing frequency. We do note that the X-ray and radio observations were not performed simultaneously but merely as close as available, and the final observation (epoch 10) shows that the radio variability can even change within two consecutive segments of $40$ minutes. 

\begin{table*}
	\centering
	\caption{Summary of variability within the X-ray and VLA radio observations. We list the average, minimum and maximum value of the inferred $6$ GHz flux density per epoch, taking both variations in low band flux density and spectral index into account when calculating these values. As an estimate of the radio variability per epoch, we also show the fractional variability $F_{\rm var}$ at both the low (5.5/4.5 GHz for ATCA/VLA) and high (9.0/7.5 GHz) frequency observing band per epoch, and for the background source (`\textit{background}'). Finally, we show the average, minimum, and maximum \textit{Swift}/XRT count rates in 100-second segments. Note that the XRT observations used for epoch 1, 2, and 3 were taken in WT mode, and for epochs 1 and 2 was only 30 seconds long.}
	\label{tab:variability}
	\begin{tabular}{lccccccccccc} %
 & \multicolumn{3}{c}{Inferred $6$ GHz radio flux densities [$\mu$Jy]} & & \multicolumn{3}{c}{Fractional radio variability} & & \multicolumn{3}{c}{\textit{Swift}/XRT count rates [cts s$^{-1}$]} \\ \cline{2-4} \cline{6-8} \cline{10-12} 
Epoch & Average & Minimum & Maximum & & low-band & high-band & background & & Average & Minimum & Maximum \\
\hline \hline
1 & $435 \pm 16$ & $143 \pm 16$ & $874 \pm 14$ & & $37.8 \pm 0.5$  & $44.2 \pm 0.3$  & $19.0 \pm 0.8$ & & $0.26 \pm 0.01 $ & -- & -- \\
2 & $557 \pm 32$ & $453 \pm 46$ & $663 \pm 93$ & & $23.9 \pm 2.5$\% & $16.7 \pm 2.2$\% & $17.7 \pm 3.7$\% & & $0.26 \pm 0.01 $ & -- & -- \\
3 & $250 \pm 16$ & $117 \pm 85$ & $280 \pm 35$ & & $33.7 \pm 5.2$\% & $21.8 \pm 2.8$\% & $9.0 \pm 7.2$\% & & $0.10 \pm 0.03 $ & $ 0.024 \pm 0.01$ & $0.18 \pm 0.05$ \\
4 & $282 \pm 15$ & $180 \pm 37$ & $329 \pm 45$ & & $28.3 \pm 3.4$\% & $16.4 \pm 4.8$\% & $14.8 \pm 4.0$\% & & $0.07 \pm 0.01 $ & $0.013^{+0.019}_{-0.009}$ & $0.37^{+0.55}_{-0.27}$ \\
5 & $365 \pm 27$ & $229 \pm 61$ & $547 \pm 41$ & & $29.2 \pm 3.6$\% & $27.0 \pm 1.6$\% & $15.3 \pm 2.8$\% & & $0.22 \pm 0.01 $ & $0.058^{+0.032}_{-0.024}$ & $0.68 \pm 0.13$ \\
6 & $245 \pm 24$ & $57 \pm 50$ & $387 \pm 44$ & & $49.0 \pm 5.3$\% & $32.5 \pm 3.8$\% & $15.3 \pm 4.4$\% & & $3.19 \pm 0.11 $ & $0.06^{+0.03}_{-0.02}$ & $13.8 \pm 1.3$ \\
7 & $204 \pm 21$ & $105 \pm 86$ & $291 \pm 45$ & & $33.7 \pm 5.1$\% & $22.7 \pm 4.2$\% & $8.9 \pm 4.3$\% & & $3.19 \pm 0.11 $ & $0.06^{+0.03}_{-0.02}$ & $13.8 \pm 1.3$  \\
8 & $185 \pm 16$ & $ 80 \pm 44 $ & $285 \pm 73$ & & $25.9 \pm 6.4$\% & $57.7 \pm 3.8$\% & $12.4 \pm 3.2$\% & & $0.16 \pm 0.02 $ & $0.07^{+0.04}_{-0.03}$ & $0.51 \pm 0.08$  \\
9 & $157 \pm 19$ & $121 \pm 48$ & $182 \pm 47$ & & $17.0 \pm 8.7$\% & -- & $8.1 \pm 3.0$\% & & $0.16 \pm 0.02 $ & $0.03 \pm 0.02$ & $0.23 \pm 0.05$ \\
10 & $171\pm11$ & $132 \pm 47$ & $300 \pm 42$ & & $23.1 \pm 3.5$\% & $32.3 \pm 2.3$\% & $19.0 \pm 2.2$\% & & $0.28 \pm 0.01$ & $0.04 \pm 0.02$ & $1.17 \pm 0.16$ \\
\hline
	\end{tabular}
\end{table*}

\subsection{The X-ray -- radio luminosity plane}
\label{sec:results_lxlr}

Finally, we turn to the X-ray -- radio luminosity plane of hard-state X-ray binaries. While Sw J1858 did not reside in a classical hard state, this figure can offer a comparison with jets from other X-ray binaries, especially in terms of radio luminosity. As these radio luminosities are plotted at $6$ GHz in our comparison sample, we calculated $6$-GHz flux densities using the $4.5$ GHz flux density and the spectral index for the VLA observations. We propagated the errors on both and list the resulting luminosities in Table \ref{tab:variability}. In Figure \ref{fig:lxlr}, we show the resulting X-ray -- radio luminosity plane, showing Sw J1858 alongside the sample collected in \citet{gallo2018}\footnote{The comparison sample from \citet{gallo2018} can also be found on  \href{https://jakobvdeijnden.wordpress.com/radioxray/}{https://jakobvdeijnden.wordpress.com/radioxray/}}. The assumed distance of $15$ kpc \citep{buisson20b} might be overestimated due to the effects of obscuration by local material, which would move Sw J1858 slightly in the bottom left direction. 

Compared to other neutron star systems, Sw J1858 is radio brighter: it overlaps more with the black hole sample, that is systematically radio brighter than the neutron star sample at a given X-ray luminosity \citet{migliari2006,gallo2018}. However, this might be an effect of strong local absorption in the system: as we will discuss in Sections \ref{sec:discuss_xrays} and \ref{sec:eddington}, such absorption can be invoked to explain the strong optical and X-ray flaring. In the X-ray -- radio luminosity plane, intrinsic absorption moves the source to lower X-ray luminosities, making it appear more radio bright compared to other sources at those luminosities. Hints for this effect might be visible in the two X-ray brightest points, which show a similar radio luminosity as the X-ray fainter observations and are therefore more consistent with other neutron star observations. Alternatively, radio luminosity differences between sources might arise from distinct boosting factors of jets viewed at different inclinations \citep{motta2018}. However, the two X-ray brightest points argue against such a scenario in Sw J1858, as these two radio observations are not significantly radio-brighter than the neutron star sample.

From the figure, no clear correlation between the X-ray and radio luminosity can be distinguished: the radio luminosities remain similar despite a jump in \textit{observed} X-ray luminosity by a factor $\sim 10$ in epochs 7 and 8. However, several effects could mask an X-ray -- radio coupling, if it were present. Firstly, if local absorption not only reduces the X-ray luminosity but also changes between epochs, the shape of any correlation between X-ray and radio will also be altered. Secondly, the X-ray and radio observations were not taken simultaneously. Given the strong variability observed \textit{between} \textit{Swift} observations, the source likely changed in X-ray luminosity significantly during the time separating the VLA and \textit{Swift} observations. Finally, both the X-rays and radio show significant variability \textit{during} the observations, which is not reflected by plotting only the average X-ray and radio luminosities. 

To quantify the range in radio luminosity spanned by each epoch, we calculated the inferred $6$ GHz flux density in every $3$-minute time bin, and selected the minimum and maximum value. In X-rays, where we could not obtain sufficient spectral information on short time scales, we consider count rates\footnote{In other words, we effectively assume a constant spectral shape during each \textit{Swift} observation, independent of flux or flaring. We note that this is an oversimplification, as the spectral shape could change during flares.}: for each epoch, we selected the average count rate, and the minimum and maximum count rates on a time scale of 100 seconds. We then estimated the range in X-ray luminosity by multiplying the average luminosity by the ratios of minimum versus average and maximum versus average count rates. Finally, we plotted the results of this exercise in Figure \ref{fig:lxlr_var}. Here, the error bars reflect the range in X-ray and radio luminosity spanned during the epoch. Taking this range into account, the epochs now overlap in both X-ray and radio luminosity, showing that the variability washes out any correlations or coupling that might be observed in strictly simultaneous observations.

Finally, we stress that the high radio to X-ray luminosity ratio and lack of observed luminosity coupling discussed above, stand out mostly in comparison to hard state X-ray binaries. The radio jet of Sw J1858 appears to be of the compact type, determined from its inverted ($\alpha > 0$) observation-averaged spectral index, as seen in the other hard state sources as well. However, from an X-ray point of view, Sw J1858 did not reside in a classical hard state. Therefore, in addition to the effects of variability and obscuration discussed above, the difference in X-ray state also likely contributes to the deviant behaviour of Sw J1858. We discuss this in more detail in Section \ref{sec:eddington}.

\section{Discussion}
\label{sec:discuss}

\subsection{X-ray spectral modelling}
\label{sec:discuss_xrays}
To determine the \textit{observed} X-ray flux around the time of the radio observations, we fitted the closest \textit{Swift}/XRT spectrum in time with a phenomenological power law model. For several reasons, this model does not provide a full physical description of the X-ray spectrum; firstly, we fix the interstellar hydrogen column density to zero, as the fitted value of the absorption pegged at this lower limit. Secondly, the source is highly variable in X-rays, with strong spectral changes during flares \citep[][]{hare2020}. Finally, as noted first by \citet{reynolds2018}, the spectral shape signals significant local absorption with a more complicated underlying spectral shape, similar to V404 Cyg \citep{motta2017} and V4641 Sgr \citep{maitra2006,morningstar2014}. 

One of the Swift/XRT observations fitted in this work (ObsID 10955004; epoch 3) was analysed by \citet{reynolds2018} with a more complicated partial covering model: \textsc{phabs * (bapec + pcfabs*po)}. They measure a 0.5--10 keV flux, corrected for ISM absorption but \textit{not} local absorption by the partial covering, consistent with the power law estimate. The intrinsic power law flux, corrected for partial covering, is however a factor $\sim 4$ higher. To extend this comparison between the two models, we also fitted the brightest (both in flux and total counts) Swift/XRT epochs (ObsId 10955008) with the partial covering model. Again, we find consistent values for the locally-absorbed flux (i.e. only ISM corrected), while the intrinsic power law flux is a factor $\sim 2.3$ higher (i.e. after correcting for local absorption) in the 0.5--10 keV range. 

The low flux of Sw J1858 prevents us from a systematic analysis of all Swift/XRT spectra with the more complicated intrinsic absorption model of \citet{reynolds2018} -- when trying such an analysis, we find poorly constrained and highly degenerate spectral parameters. However, the above considerations show that using a non-physical power law model results in similar measurements of the \textit{observed} fluxes. Similarly, we find that local absorption only alters the measured flux by a factor of roughly $2$--$4$, which is within the systematic uncertainty introduced by the X-ray flaring (see e.g. Figure \ref{fig:lxlr_var}). We will discuss the effects of local absorption in more detail in Section \ref{sec:eddington}. 

\subsection{The origin of radio variability of Sw J1858}
\label{sec:origininSwJ1858}

In this section, we will discuss what might cause the observed radio variability in Sw J1858. Firstly, could it be caused by changes in the geometry of the jet, such as precession, leading to a variable jet boosting factor? Jet precession has been directly observed \citep{coriat2019,millerjones2019} or inferred \citep[e.g.][]{gallo2014} in several LMXBs, and has been proposed as the mechanism behind quasi-periodic oscillations in X-rays at (sub-)second timescales \citep{stella1998,ingram2009,ingram2016fullmod}. However, we do not find any evidence for precession in our observations: most of the radio variability appears to be stochastic, and the exceptions, such as the radio flare in epoch 5, do not show (quasi)-periodic variations. Therefore, we rule out such a scenario. 

Secondly, we can consider whether radio intensity scintillation caused by scattering in the interstellar medium might contribute to the observed variations in Sw J1858. Following \citet{Pandey2006}, we use the online \textsc{NE2001} tool\footnote{\href{https://www.nrl.navy.mil/rsd/RORF/ne2001/}{https://www.nrl.navy.mil/rsd/RORF/ne2001/}} by \citet{cordes2001} to calculate the scintillation characteristics in the direction of Sw J1858, for several distances. For all observing freqencies, the predicted time scales for amplitude variations due to scintillation are shorter than our time resolution. The affected angular size scales are of the order of $50$--$100$ $\mu$as. At 15 kpc, this size translates to a scale of $5$--$10\times10^{7}$ gravitational radii for a $1.4$~$M_{\odot}$ neutron star. It is (particularly for neutron star X-ray binaries) poorly known from what physical distances from the accretor the cm-wavelength radio emission originates. For black holes, the recent correlated X-ray and radio timing study of Cyg X-1 suggests similar or larger scales than those affected by scintillation \citep{tetarenko2019}. However, due to the mismatch in time scales, we consider it unlikely that scintillation contributes significantly to the observed variability. 

Thirdly, the radio variability does not appear to be caused by the launch of discrete ejecta. The launch of such ejecta has been associated with strong radio variability as observed in both V404 Cyg \citep{tetarenko2017_v404} and V4641 Sgr \citep{rupen2003b,rupen2003,rupen2004,rupen2004b} during states of extreme X-ray flaring, similar to that seen in Sw J1858. In Sw J1858, we do not observe the steady, optically-thin spectra associated with a bright (or decaying) radio flare that are expected in this scenario; indeed, during the flares observed in epochs 1 and 5, the radio spectrum is flat or inverted during the peak. Also, we do not see a smoothing of the variability towards lower frequencies, as observed in this scenario \citep[e.g.,][]{tetarenko2017_v404}. Similarly, we find no evidence that the radio variability is directly related to the contemporaneous level of X-ray flaring. While the radio and X-ray observations are not strictly simultaneous, we do not find a relation between the fractional variability in X-rays and radio. This direct relation between the X-ray and radio flaring was neither observed in V404 Cyg \citep{tetarenko2017_v404}. This is consistent with a scenario where the X-ray flaring is caused by strong obscuration of the inner accretion flow combined with a high mass accretion rate (See Section \ref{sec:eddington}), as the obscuration would not affect the jet. 

Alternatively, the radio variability might be caused by intrinsic variability in the accretion flow and mass accretion rate propagating into the jet \citep[e.g.,][]{malzac2014}. This scenario can explain why the radio spectrum remains, on average, inverted over time, as the compact jet does not change its global morphology. We observe that the variability is stronger, in most observations, at the lower observing frequency. This difference in variability could arise if the fluctuations are smeared out to longer time scales further down the jet, at lower emission frequencies, therefore better matching the three to five minute time scales probed by our analysis. Using simultaneous X-ray and radio observations, \citet{tetarenko2019} showed that a similar scenario is at play in Cyg X-1. However, without simultaneous X-ray coverage of, for instance, the flare in epoch 5, it is challenging to directly show how the intrinsic variability propagates from the accretion flow to the jet. 

The variability in the accretion flow could arise from, for instance, the mass accretion rate fluctuations in the disk that are typically used to explain rapid broad-band X-ray variability  \citep{ingram2011,rapisarda2016}. Alternatively, disk tearing due to Lense-Thirring precession, caused by a misalignment between the accretor and disk spin axis, was proposed by \citet{nixon12a} and \citet{nixon12b} to cause variations in the accretion rate. To test whether the latter scenario could contribute, we can estimate the minimum required spin misalignment for Sw J1858. Neutron stars in LMXBs have typical dimensionless spin parameters $a$ between $\sim 0.05$ and $0.25$, for a range of equations of state \citep{haensel09,lo10}. We assume a standard viscous disk with $\alpha=0.1$ and a moderately vertically extended flow ($H/R \approx 0.1$), valid if the Sw J1858 is intrinsically accreting at higher rates than observed (see Section \ref{sec:eddington}). Then, from equation 9 in \citet{nixon12b}, we find that the spins must be misaligned by at least $8.6\degree$ ($1.7\degree$) for $a=0.25$ ($a=0.05$). If we add a second requirement, namely that the disk must tear at a radius larger than the neutron star radius (assumed to be $10$ km), we find that the misalignment angle must be $>14\degree$ for $a=0.25$, and even larger for smaller spins. The detection of Type-I bursts in Sw J1858 suggest a relatively weak magnetic field; therefore, such a spin misalignment should have persisted while the magnetic field decayed, in order for this scenario to be feasible. 

\subsection{Sw J1858 in the $L_X$--$L_R$ diagram: the effects of variability}
\label{sec:lxlr_discussion}

Here, we briefly turn to the behaviour of Sw J1858 in the X-ray -- radio luminosity plane and the effects of X-ray and radio variability. We first re-emphasize that Sw J1858 was not observed in the canonical hard X-ray state, where the other sources in the $L_X$--$L_R$ diagram reside. Therefore, it is not surprising that we do not observe a clear coupling between the X-ray and radio luminosity: such a coupling is also not observed during flaring states of the black holes V404 Cyg \citep{tetarenko2017_v404,tetarenko2019_v404} or GRS 1915+105 \citep{pooley1997}. Similarly, one cannot easily state whether Sw J1858 is underluminous compared to the sample, even if it intrinsically accretes at higher rates (see Section \ref{sec:eddington}).  

Recently, \citet{plotkin2019} analysed all quiescent observations of V404 Cyg, focusing on the variability properties in the radio and X-ray band. They showed that, while the source shows a correlation between its X-ray and radio luminosity during the non-flaring states, this correlation disappears when accounting for intra-observational variability and non-simultaneity between X-ray and radio observations. We similarly find that, when plotting the full range in X-ray and radio luminosities reached in an observation, different epochs overlap in the $L_X$--$L_R$ diagram. So while we do not observe an X-ray -- radio coupling during the time-averaged observations, any such coupling detected would not have been reliable given the level of observed variability. 

\subsection{A radio comparison between Sw J1858 and Eddington-limited LMXBs}
\label{sec:eddington}

In the black hole LMXBs V404 Cyg and V4641 Sgr, the observed strong optical and X-ray flaring can be explained by an extreme, $\sim$Eddington mass accretion rate, combined with a high inclination \citep[][e.g.]{wijnands2000,munozdarias2016,motta2017,sanchezfernandez2017_v404}. Due to the high mass accretion rate, the inner accretion flow puffs up into a vertically extended flow, launching a highly-ionized and clumpy outflow. As a result, the inner accretion flow is typically blocked from view, reducing the observed X-ray flux greatly, with brief intervals providing an unobscured view causing the large flares. As similarly strong flaring is observed in the X-ray and optical bands in the neutron star Sw J1858 \citep{ludlam2018,hare2019,vasilopoulos2018,baglio2018,rajwade2018,rajwade2019,paice2018,hare2020}, a similar scenario might be at play in the accretion flow there \citep[although see][for a recent discussion on possible differences between the X-ray behaviour of V404 Cyg, V4641 Sgr, and Sw J1858]{hare2020}. Do the radio (variability) properties of Sw J1858 fit with such high-inclination, highly-obscured, Eddington-limited accretion? 

The recent report by \citet{buisson20b} of the detection of Type-I bursts in Sw J1858 shows the presence of a neutron star primary. Therefore, here, we first consider whether Eddington-limited accretion in such a neutron star LMXB might explain the observed radio behaviour in Sw J1858. The neutron star LMXBs with the highest mass accretion rates are the Z-sources, which are thought to accrete near or at the Eddington luminosity, tracing out Z-shaped tracks in their X-ray color-color diagrams \citep{hasinger1989, homan2010}. Z-sources can show strong changes in radio brightness, related to the branch they are positioned on in their color-color diagram track; they are radio brighter and more variable in the Horizontal Branch than in the Flaring and Normal Branches \citep{penninx1988,hjellming1990_scox1,hjellming1990_cygx2,spencer2013,motta2019}. Time-resolved radio studies of the different branches in Sco X-1 by \cite{hjellming1990_scox1} and Cyg X-2 by \citet{hjellming1990_cygx2} show that these sources have similar levels of radio variability and luminosity to Sw J1858 during their radio-faint Flaring and lower Normal Branches. However, in the X-ray -- radio luminosity plane, these sources are located to the right of Sw J1858, around the neutron star Eddington limit of $2\times10^{38}$ erg/s. For Sw J1858 to be similar to these sources, it would therefore have to be viewed a high inclination, reducing its observed X-ray flux and masking the Z-source variability properties through obscuration. However, a clear radio difference between Sw J1858 and the Z-sources is the radio spectral index: contrary to Sw J1858, Z-sources typically show steep spectra and are indeed associated with the launch of (resolved) discrete ejecta \citep{motta2019}. Also, in this scenario, Sw J1858 should have not have resided in the much more radio bright Horizontal Branch during any of the observations, which is unlikely given the time scales and commonness of transitions between the branches \citep{homan2010}. 

We can alternatively make the comparison with the black hole V404 Cyg, which was extensively monitored in the X-ray and radio bands during its 2015 outburst. At the peak of its Eddington-limited outburst, it showed a strongly-flaring X-ray state \citep{motta2017,sanchezfernandez2017_v404} in which the observed radio variability -- flaring up to flux densities of $1$ Jy -- is associated with the launch of discrete ejecta \citep{tetarenko2017_v404}. As it decayed towards quiescence, V404 Cyg transitioned into a more calm state where a compact, but still variable (albeit at a lower level) radio jet starts to dominate the radio emission at the level of several mJy \citep{tetarenko2019_v404}. In addition, in the first several days of the outburst, before the peak, the X-ray band already showed flaring while AMI-LA radio monitoring reveals a relative faint and variable radio source at tens to hundreds of mJy \citep{munozdarias2016}.

While the peak of the V404 Cyg outburst might resemble the X-ray behaviour of Sw J1858 most closely, the radio properties of both sources do not match in this state: Sw J1858 does not show evidence for optically thin, discrete ejecta, and is significantly underluminous compared to V404 Cyg in this state. In terms of levels of radio variability, radio spectral index, and radio luminosity, Sw J1858 is in fact more similar to the behaviour of V404 Cyg during the decay of its 2015 outburst: a large distance of $15$ kpc to Sw J1858 can account for the difference in radio flux density, and both sources show an inverted spectrum. In addition, \citet{tetarenko2019_v404} suggest that the radio variability observed in this state have propagated down the jet from the accretion flow, as we suggest for Sw J1858. However, during this state, the X-ray behaviour does not match: at that point, the X-ray flaring had halted in V404 Cyg. That leaves the very start of the outburst. Again, for the Type-I burst distance to Sw J1858 of $15$ kpc, the radio luminosities are similar, while V404 Cyg already showed X-ray flaring at this point \citep{munozdarias2016}. Comparing the radio variability properties is however difficult, as the detailed variability studies by \citet{tetarenko2017_v404} and \citet{tetarenko2019_v404} of V404 Cyg do not include the AMI-LA observations from this part of the outburst. Also, a clear difference in this scenario is the duration: this short lived state in the 2015 outburst of V404 Cyg was followed by the radio-flaring outburst peak only days later, while in Sw J1858, the same behaviour has persisted for months (see Figure \ref{fig:longtermlc}). 

\section{Conclusions}

In this paper, we have reported radio monitoring of the new transient neutron star X-ray binary Sw J1858. While the source does not show a canonical hard state, we detect a compact jet in every observation. Time-resolving the observations, we find that the jet emission is highly variable in most observations. Finally, we do not observe a clear correlation between the radio and X-ray luminosity of Sw J1858, as also seen in other LMXBs showing similar flaring X-ray states. We find that the radio variability is most likely associated with a compact jet responding to variations in the accretion flow propagating down the jet. Precession, scintillation, and the launch of discrete ejecta can be ruled out as possible explanations. Finally, we discuss how the radio properties of Sw J1858 compare to other LMXBs showing similar X-ray flaring behaviour, finding that none of these previously observed sources provides a perfect match. Therefore, while several of the radio properties of Sw J1858 are similar to either Z-sources, V404 Cyg or V4641 Sgr, it also shows as-of-yet unseen behaviour. 

\section*{Data Availability Statement}
The data underlying this article are available in Zenodo, at \href{https://dx.doi.org/10.5281/zenodo.3878127}{https://dx.doi.org/10.5281/zenodo.3878127}.

\section*{Acknowledgements}
The authors thank the VLA director and schedulers for approving, rapidly scheduling, and coordinating the DDT observation included in this study. This research made use of Astropy,\footnote{http://www.astropy.org} a community-developed core Python package for Astronomy \citep{astropy2013, astropy2018}. JvdE and ND are supported by an NWO Vidi grant awarded to ND. TDR is supported by a NWO Veni grant. DA and DJKB acknowledge support from the Royal Society. MAP is funded by the Juan de la Cierva Fellowship IJCI--2016-30867. TMD is funded by the Ram\'on y Cajal Fellowship RYC-2015-18148. MAP and TMD acknowledge support by the Spanish MINECO grant AYA2017-83216-P. FF and DD acknowledges support from the Royal Society International Exchanges "The first step for High-Energy Astrophysics relations between Argentina and UK". MOA acknowledges support from the Royal Society through Newton International Fellowship program. GRS acknowledges support from an NSERC Discovery Grant (RGPIN-2016-06569). This research has made use of data and software provided by the High Energy Astrophysics Science Archive Research Center (HEASARC) and NASA's Astrophysics Data System Bibliographic Services. The Australia Telescope Compact Array (ATCA) is part of the Australia Telescope, which is funded by the Commonwealth of Australia for operation as a National Facility managed by CSIRO. The National Radio Astronomy Observatory is a facility of the National Science Foundation operated under cooperative agreement by Associated Universities, Inc.


\label{lastpage}

\end{document}